\DeclareUnicodeCharacter{2212}{\textendash}
\pdfoutput=1 
\documentclass[prl,twocolumn,showpacs,graphicx,print]{revtex4-1}
\usepackage{graphicx}
\usepackage{dcolumn}
\usepackage{bm}
\usepackage{epsfig}
\usepackage{lipsum}
\usepackage{natbib}
\usepackage{ulem,cancel,soul}
\usepackage[fleqn]{amsmath}
\usepackage{slashed}
\usepackage{mathrsfs}
\usepackage{amssymb}
\usepackage{times}
\usepackage{psfrag}
\usepackage[english]{babel}
\usepackage{multirow}
\usepackage{color}
\begin{document}

\title{Alcubierre Warp Drive in Bohmian Quantum Gravity}

\author{Sijo K. Joseph}
\affiliation{GITAM Deemed to be University, Hyderabad, Telangana, India.}
\email{skizhakk@gitam.edu}

\begin{abstract}
Alcubierre warp drive metric is coupled to quantum mechanical scalar matter field. The requirement of the exotic matter for the warp drive is mapped into a conformal wave equation. This result into a  fourth order partial differential equation in terms of the quantum mechanical  density.  Finding a proper quantum mechanical  density obeying the proposed partial differential equation will be a resolution to the exotic matter problem of Alcubierre warp drive in  Bohmian Quantum Gravity context.
\end{abstract}

\date{\today}


\maketitle

\section{Introduction}
Generalization of Einstein's theory of gravity is under intense theoretical exploration. Main motivation behind this research involves
 both mathematical and physical reasons. In a purely mathematical perspective, the Lagrangian density $\mathscr{L}$ of Einstein's
  gravity is simply linear to the Ricci scalar $R$. One is always tempted to generalize this simple Einstein-Hilbert Lagrangian to more
  general ones, resulting into  $f(R)$ theories~\cite{BeyondEinsteinGravity,Bergmann1968,fR_HamFormlation,Nojiri_fofr_intro,EnerfofrandBransDicke} and $f(R,T)$ theories etc.~\cite{FofRT1} . In addition, there
   are several physical motivations to study generalized Einstein theory, for example, the dark energy problem is one among them
   ~\cite{Capozziello2006}. Quantum
  gravity considerations is also there, which hints the extension of Einstein's theory for a concise physical picture.

Even though there are several mathematical route for the generalization of Einstein's theory, $f(R)$ theory is mathematically more
 straightforward. Later it was shown that $f(R)$ theory of gravity~
\cite{fofr_agravity,Capozziello2011} is equivalent to
 scalar tensor theory of gravity~\cite{Ntahompagaze2017}. Based on the nature of the covariant derivative of the metric tensor,
 Einstein's gravity can also be generalized in a different manner. Metric compatibility ($\nabla_{\sigma}g_{\mu\nu}=0$) and zero torsion ($\Gamma^{\sigma}_{\mu\nu}-\Gamma^{\sigma}_{\nu\mu}=0$) are the two essential conditions in Einstein's theory. Both of these conditions can be relaxed and different kind of extensions Einstein's theory can be obtained. Relaxing the metric compatibility condition, one is naturally lead to Weyl differential geometry~\cite{WeylReview2017}. Due to the research in the last few decades, it is identified that the Weyl type of differential geometry is very useful to incorporate quantum mechanical scalar matter fields as a correction to Einstein's theory~\cite{Shojai_Article} . Based on the deBroglie-Bohm version of quantum theory~\cite{Bohm1975,BohmI,BohmII}, one is able to couple quantum theory with classical gravity. There are several research work along this interesting direction. 

Usual quantum gravity approaches like Loop Quantum Gravity focuses on the quantization of gravity~\cite{AshtekarBook2017,rovelli_vidotto_2014} while  Bohmian Quantum Gravity gives only a geometrical way to couple quantum mechanical scalar matter fields with Einstein's gravity~\cite{Shojai_Article,Shojai_ScalarTensor,ShojaiPRD99}. Note that  gravity is still treated as classical  and we are only focusing on how classical field theories are coupled each other. We will first focus on the essential equations in Bohmian Quantum Gravity and more specifically the  stress-energy tensor is analyzed in connection with warp drive. Then we will explore Alcubierre warp drive metric, taking it as  the background space-time  coupled to quantum mechanical scalar matter field. Here we search for the properties of the quantum system in order to support  Alcubierre metric as the background space-time.
\section {Bohmian Quantum Gravity}
In previous research works~\cite{GabayJoseph1,SKJoseph2020,SKJoseph2021}, the following action is taken into account, where the quantum mechanical scalar matter field is coupled to gravity in a geometric manner (de Broglie-Bohm manner),
\begin{eqnarray}
& & A[g_{\mu\nu},{\Omega}, S, \rho, \lambda]=\nonumber\\
& & \frac{1}{2\kappa}\int{d^4x\sqrt{-g}\left(R\Omega^2-6\nabla_{\mu}\Omega\nabla^{\mu}\Omega\right)}  \nonumber \\
& & +\int{d^4x\sqrt{-g} \left(\frac{\rho}{m}\Omega^2 \nabla_{\mu}S \nabla^{\mu}S-m\rho\Omega^4\right)} \nonumber \\
& & +\int{d^4x\sqrt{-g}\lambda\left[\ln{\Omega^2}-\left(\frac{\hbar^2}{m^2}\frac{\nabla_{\mu}\nabla^{\mu}\sqrt{\rho}}{\sqrt{\rho}}
\right)\right]} \label{Actioneq}. 
\end{eqnarray}
It is already shown that, minimizing the action with respect to $\rho$ and $S$ 
leads to real and imaginary parts of the Generalized Klein-Gordon Equation. 
Here Eq.~\ref{EqMotion}  describes the real part of the Klein-Gordon Equation and it is given by, 
\begin{equation}
\begin{split}
\nabla_{\mu}S \nabla^{\mu}S-m^2\Omega^2+\frac{\hbar^2}{2m 
\Omega^2\sqrt{\rho}}\Bigl[\Box{\Bigl(\frac{\lambda}{\sqrt{\rho}}\Bigr)}-\lambda\frac{\Box\sqrt{\rho}}{\rho}\Bigr]=0. \label{EqMotion} 
\end{split}
\end{equation}
Here Eq.~\ref{ContiEqn} gives the generalized continuity equation which is the imaginary part of the more general Klein-Gordon Equation (see Ref~.\cite{GabayJoseph2} for more details),
\begin{eqnarray}
\nabla_{\mu}(\rho\Omega^2\nabla^{\mu}S)=0 \label{ContiEqn}.
\end{eqnarray}
One can also see that  the constraint equation is given by,
\begin{eqnarray}
\Omega^2=\exp{\left(\frac{\hbar^2}{m^2}\frac{\nabla_{\mu}\nabla^{\mu}\sqrt{\rho}}{\sqrt{\rho}}\right)}. \label{CnstrEq}
\end{eqnarray} 
where the conformal factor is identified as a quantum mechanical quantity i.e.
 the exponential of the Bohmian quantum potential. 
In varying the action with respect to $\Omega$, one obtains an equation of motion for the scalar curvature $R$ 
\begin{eqnarray}
R\Omega+6 \Box\Omega +\frac{2\kappa}{m}\rho \Omega (\nabla_{\mu}S \nabla^{\mu}S-2m^2\Omega^2)+\frac{2\kappa\lambda}{\Omega}=0. \nonumber \\
\label{TraceEq}
\end{eqnarray}
Finally, varying the action with respect to the inverse metric tensor ${g}^{\mu\nu}$, the conformally transformed Einstein equation, along with its stress-energy tensors, are obtained
\begin{eqnarray}
\mathcal{G}_{\mu \nu}=T^{\bf{matter}}_{\mu\nu}(S,\rho)+T^{\bf{qm}}_{\mu\nu}(\Omega)+T^{\bf{med}}_{\mu\nu}(\lambda,\rho) 
\label{EinsteinEq}
\end{eqnarray}
where the stress-energy tensors are given by,
\begin{eqnarray}
T^{\bf{matter}}_{\mu\nu}(S,\rho)&=& -\frac{2\kappa}{m}\rho \nabla_{\mu}S \nabla_{\nu}S 
 +\frac{\kappa}{m} \rho\,g_{\mu\nu}\nabla_{\sigma}S \nabla^{\sigma}S \nonumber \\
 & &-\kappa m \rho \Omega^2 g_{\mu\nu}. \label{tuv_matter}
\end{eqnarray}
\begin{eqnarray}
T^{\bf{qm}}_{\mu\nu}(\Omega) &=&\frac{(g_{\mu \nu}\Box{\Omega^2}- 
\nabla_{\mu}\nabla_{\nu}{\Omega^2})}{\Omega^2}+6\frac{\nabla_{\mu}\Omega 
\nabla_{\nu}\Omega}{\Omega^2} \nonumber \\
& & -3 g_{\mu \nu} \frac{\nabla_{\sigma}\Omega \nabla^{\sigma}\Omega}{\Omega^2}. \label{tuv_qm}
\end{eqnarray}
\begin{eqnarray}
T^{\bf{med}}_{\mu\nu}(\lambda,\rho)= 
-\frac{\kappa\hbar^2}{m^2\Omega^2}\Biggl[\nabla_{(\mu}&&\sqrt{\rho}\nabla_{\nu)}\Bigl(\frac{\lambda}{\sqrt{\rho}}\Bigr) \nonumber \\ 
&&-g_{\mu\nu}\nabla_{\sigma}\Bigl(\lambda\frac{\nabla^{\sigma}{\sqrt{\rho}}}{\sqrt{\rho}}\Bigr)\Biggr]. \label{TVacuum}
\end{eqnarray}
 Note that, the resulting Einstein's equation is composed of stress-energy tensors related to the matter $T^{\bf{matter}}_{\mu\nu}(S,\rho)$, quantum $T^{\bf{qm}}_{\mu\nu}(\Omega)$ and mediating field $T^{\bf{med}}_{\mu\nu}(\lambda,\rho)$ contributions.
 The vacuum density $\lambda$ obeys a first order differential equation coupled to  $\sqrt{\rho}$ , which is given by,
\begin{eqnarray}
\lambda=\frac{\hbar^2}{m^2(1-Q)}\nabla_{\mu}\Bigl(\lambda\frac{\nabla^{\mu}\sqrt{\rho}}{\sqrt{\rho}} \Bigr). \label{LambdaNiceEq1}
\end{eqnarray}
This $\lambda$ equation is obtained by comparing trace equation given in Eq.~\ref{TraceEq}  with the trace of the tensor equation given in Eq.~\ref{EinsteinEq} . It is found that  $\lambda$ term can give additional corrections to standard quantum theory (see Ref.~\cite{SKJoseph2020,SKJoseph2021} for more details) .

\section{Alcubierre Warp Drive}
Usually Alcubierre metric is studied using Einstein's General Theory of Relativity and one is  encountered with the problem of the exotic matter~\cite{Alcubierre_1994}. Later, considerable amount of research works were carried out in relation to the physical feasibility of the  Warp drive solution~\cite{VanDenBroek2000}, its energy requirements~\cite{VanDenBroek1999,Matt2022}. Advacing further, some of the modifications to Alcubierre warp drive were also suggested~\cite{Natario2002,White2003,White2011}. Recently  various matter-energy sources is explored in order to see whether it  supports Alcubierre warp drive metric solution and a relation to  Burgers' equation  is found~\cite{OLSPereira2020_1,OLSPereira2021_1, OLSPereira2021_2,OLSPereira2021_3}.

Here we focus on the more generalized theory of Einstein's gravity, i. e. the Bohmian Quantum Gravity which incorporate quantum mechanical matter field coupling to gravity using conformal factor. In order to explore the warp drive in the context of Bohmian Quantum Gravity, we use the following assumption that the background space-time is Alcubierre warp-drive metric
which is coupled to a quantum mechanical scalar matter field $\phi=\sqrt{\rho}\exp{(\frac{i}{\hbar}S)}$. In Bohmian Quantum Gravity, this complex scalar matter-field can be coupled to gravity and the total metric of the system is given by $g_{\mu\nu}^{\bf{qm+gr}}=\Omega^2\,g_{\mu\nu}^{\bf{gr}}$ where $\Omega^2$ is the quantum mechanical conformal factor. Here the gravitational background metric $g_{\mu\nu}^{\bf{gr}}$ is simply represented as ${g}_{\mu\nu}$.
In $(+ - - -)$ signature, Alcubierre metric can be written as, 
\begin{eqnarray}
ds^2&=&g_{\mu\nu}dx^{\mu}dx^{\nu} \nonumber \\
           &=&{\bigl(1-v_{s}(t)^2f(r_{s}(t)^2\bigr)}{dt^2}\nonumber \\& & +2v_{s}(t)^2f(r_{s}(t)) {dx}\,{dt} -dx^2-dy^2-dz^2,
\end{eqnarray}
where $v_{s}(t)=\frac{dx_{s}(t)}{dt}$ is the warp bubble velocity.  While the metric tensor $g_{\mu\nu}$ with $(+ - - -)$ signature can be written as,

\begin{eqnarray}
g_{\mu\nu}=
\begin{pmatrix} 
	 1 - v_{s}^2 f(r_{s})^2&v_{s} f(r_{s}) & 0 & 0 \\
	v_{s} f(r_{s}) & -1 & 0&0\\
	0 & 0 & -1 & 0 \\
	0 & 0 & 0 & -1 \\
	\end{pmatrix} \label{Alcgdown}.
 \end{eqnarray}
 
 The inverse metric tensor $g^{\mu\nu}$ can be evaluated as,
 \begin{eqnarray}
g^{\mu\nu}=
\begin{pmatrix} 
	 1 &v_{s} f(r_{s}) & 0 & 0 \\
	v_{s} f(r_{s}) & -1+v_{s}^2 f(r_{s})^2 & 0&0\\
	0 & 0 & -1 & 0 \\
	0 & 0 & 0 & -1 \\
	\end{pmatrix}\label{Alcgup}.
 \end{eqnarray}
Here $f(r_{s})$ is termed as the shape function of the warp metric since it describes the shape of the warp bubble. 
\begin{eqnarray}
f(r_{s}) =\frac{\tanh{[\sigma(r_{s} + R)]} − \tanh{[\sigma(r_{s} - R)]}}{2\tanh{(\sigma R)}},
\end{eqnarray}
where $\sigma$ is inversely related to the thickness of the warp bubble and $R$ is the warp bubble radius.
The variable $r_{s}(t)$ is defined as the distance from the center of the bubble
$[x_{s}(t), 0, 0]$ to an arbitrary point $(x, y, z)$ on the surface of the bubble and it is given by,
$r_{s}(t)=\sqrt{(x-x_{s}(t))^2+y^2+z^2}$. Alcubierre already had mentioned that this shape function can approach a step function as $\sigma\to\infty$, i.e  $|r_{s}(t)| < R$ then $f(r_{s}) = 1$, whereas for distances where $|r_{s}(t)| \gg R$ then $f(r_{s}) \to 0$.
The components for the Eulerian (normal) observers’ 4-velocities are given by,
$u^{\alpha} = [1, v_{s}(t)f(r_{s}), 0, 0]$ and $u_{\alpha} = [1, 0, 0, 0]$. Note that these vectors are considered in $(+ - - -)$ signature.
Here also one can obtain,
\begin{eqnarray}
\mathcal{G}_{\mu\nu}u^{\mu}u^{\nu}=\frac{-v_s^2 \bigl(y^2 + z^2\bigr)}{4  r_s ^2} \Bigl(\frac{df}{dr_s}\Bigr)^2.
\end{eqnarray}
Using the relation $\mathcal{G}_{\mu\nu}= {T}_{\mu\nu} $ (see Eq.~\ref{EinsteinEq}) the following expression is obtained,
\begin{eqnarray}
{T}_{\mu\nu}u^{\mu}u^{\nu}=-\frac{v_s^2 \bigl(y^2 + z^2\bigr)}{4  r_s ^2} \Bigl(\frac{df}{dr_s}\Bigr)^2. \label{ExoAmt}
\end{eqnarray}
Note that this expression is derived using the geometric arguments only. Alcubierre had started with a warp drive metric and  found a condition for the energy density in order to have a solution that he desires~\cite{Alcubierre_1994}. Then one needs to look into the proper stress-energy tensor $T_{\mu\nu}$ which can support the desired warp drive space-time structure. In usual general relativity one is compelled to accept the requirement of the exotic matter in order to the get the desired warp drive solution. Warp drive metric is no longer a solution to Einstein's General Relativity since it violate the energy conditions allowed in the theory. Hence one is instructed to look into a more generalized theory which can easily support the warp drive solution without much difficulty and Bohmian Quantum Gravity is a good candidate for that. 

Since we are taking into account the extended theory of Einstein General Relativity i.e a Scalar-Tensor Theory which contains an additional scalar degrees of freedom apart from the Einstein's gravitational tensor field. Note that the scalar field is quantum mechanical in origin.
There are many different types of scalar tensor theories but the Bohmian Quantum Gravity impose a special physical meaning to the scalar field and it is intimately related to the conformal factor in the theory which is quantum mechanical in nature. The quantum mechanical conformal factor is getting coupled to gravity. Once quantum mechanical matter field couples with gravity, one can find the exact expression for the stress energy tensor $T_{\mu\nu}$ and it contains three contributions (see Eq.~\ref{tuv_matter}, Eq.~\ref{tuv_qm} and Eq.~\ref{TVacuum}). Substituting the expression of $T_{\mu\nu}$ in the equation, one can find the condition which needs to be satisfied in the Bohmian Quantum Gravity. 

 Even-though the space-time is rigid in the Riemannian sense, it is not so in Weyl sense. 
In Riemannian geometry, the length of a vector remains constant and only the orientation changes during parallel transport . 
While in Weyl Geometry the length of a vector and orientation changes during parallel transport. 
\begin {eqnarray}
\nabla_{\alpha}g_{\mu\nu} = 0 &\implies &\text{Riemannian Geometry} \\
\nabla_{\alpha}g_{\mu\nu} = \mathcal{\sigma}_{\alpha} g_{\mu\nu} & \implies& \text{Weyl Geometry} 
\end{eqnarray}
In Riemannian geometry the covariant derivative of the metric is zero while it is not true in the Weyl Geometry. Since the length changes during the parallel transport, the covariant derivative of the metric gives a nonzero contribution. In Weyl geometry, one can think about different conformal frames where the metric is different by a conformal scaling but the physics remains the same. In Bohmian Quantum Gravity frame-work, one can see that the quantum potential $Q$ appears as the Weyl scalar field  $\mathcal{\sigma}$  since $\nabla_{\alpha}g_{\mu\nu}^{\bf{qm+gr}} = Q_{\alpha}\, g_{\mu\nu}^{\bf{gr+qm}}$ and it is purely a quantum mechanical effect. Parallel transporting a vector along a Weyl manifold will make a change in the length of the vector and such a freedom is not there in Riemannian manifold.  In other words, Space-time is rigid in the Riemannian sense but it is not so rigid in Weyl sense, hence using quantum effects one is able to manipulate space-time structure easily. This interesting fact is reflected while analyzing the contribution coming from ${T}_{\mu\nu}u^{\mu}u^{\nu}$ and major contribution is from $T^{\bf{qm}}_{\mu\nu}(\Omega)u^{\mu}u^{\nu}$. Note that the inverse gravitational constant $1/{\kappa}$ appearing as a term in front of the usual expression for the exotic matter is not present while considering  $T^{\bf{qm}}_{\mu\nu}(\Omega)u^{\mu}u^{\nu}$ term. Since we adopt the Weyl Geometry, it can be seen that the Weyl length change is associated to conformal factor $\Omega^2$ which is purely quantum mechanical in origin.  Hence quantum mechanical contribution can make space-time manipulation much easier than the usual gravitational bending of space-time.

 From our previous studies~\cite{GabayJoseph1,SKJoseph2020,SKJoseph2021}, it is found that there is a dynamical cosmological term which appears along with the $g_{\mu\nu}$ term. The dynamical cosmological term is written in terms of the quantum potential $Q$ and Lagrange multiplier $\lambda$. Hence, focusing on $T^{\bf{qm}}_{\mu\nu}(\Omega)u^{\mu}u^{\nu}$ term's $g_{\mu\nu}$ part and ignoring other contributions (they are smaller due to the presence of $\kappa$), one is lead to 

\begin{eqnarray}
{T}_{\mu\nu}u^{\mu}u^{\nu}&\approx&T^{\bf{qm}}_{\mu\nu}(\Omega)u^{\mu}u^{\nu}\nonumber \\
{T}_{\mu\nu}u^{\mu}u^{\nu}&\approx&\frac{\Box{\Omega^2}-3 {\nabla_{\sigma}\Omega \nabla^{\sigma}\Omega}}{\Omega^2}g_{\mu\nu}u^{\mu}u^{\nu}\nonumber \\
\end{eqnarray} 
Since $g_{\mu\nu}u^{\mu}u^{\nu}=1$ in $(+ - - -)$ signature, one can obtain,
\begin{eqnarray}
{T}_{\mu\nu}u^{\mu}u^{\nu}\approx\frac{\Box{\Omega^2}-3 {\nabla_{\sigma}\Omega \nabla^{\sigma}\Omega}}{\Omega^2}.
\end{eqnarray} 
This is the result from Bohmian Quantum Gravity and combining this with our previous expression for the exotic matter,  one can get,
\begin{eqnarray}
\frac{\Box{\Omega^2}-3 {\nabla_{\sigma}\Omega \nabla^{\sigma}\Omega}}{\Omega^2}=-\frac{v_s^2 \bigl(y^2 + z^2\bigr)}{4  r_s ^2} \Bigl(\frac{df}{dr_s}\Bigr)^2\label{ConWEq0}
\end{eqnarray} 
Note that, while establishing Eq.\ref{ConWEq}, one is utilizing more general  gravitational theory.
Rearranging the terms, one is lead to a Klein-Gordon type wave equation in terms of the  conformal factor $\Omega^2$, i.e
\begin{eqnarray}
{\Box{\Omega^2}-3 {\nabla_{\sigma}\Omega \nabla^{\sigma}\Omega}}+\frac{v_s^2 \bigl(y^2 + z^2\bigr)}{4  r_s ^2} \Bigl(\frac{df}{dr_s}\Bigr)^2 {\Omega^2}=0 \label{ConWEq}
\end{eqnarray} 
Let us recall the expression of exotic matter as  Alcubierre exotic function $\mbox{Alc}(x,y,z)$ and it is given by
$\mbox{Alc}(x,y,z)=\frac{v_s^2 \bigl(y^2 + z^2\bigr)}{4  r_s ^2} \Bigl(\frac{df}{dr_s}\Bigr)^2$, then Eq.~\ref{ConWEq} becomes,
\begin{eqnarray}
{\Box{\Omega^2}-3 {\nabla_{\sigma}\Omega \nabla^{\sigma}\Omega}}+\mbox{Alc}(x,y,z)\, {\Omega^2}=0 \label{ExoticWEq}
\end{eqnarray} 
Here the Alcubierre exotic function appear like the positive mass of the conformal wave equation.
Note that, the conformal wave equation is purely a quantum mechanical feature, since  $\Omega^2$ is intimately related to the quantum potential $Q$ which in turn related to the square root of the quantum mechanical matter wave density. The  constraint equation indicates that
\begin{eqnarray} 
\Omega^2=\exp{\Bigl(\frac{\hbar^2}{m^2}\frac{\Box{\sqrt{\rho}}}{\sqrt{\rho}}\Bigr)}=\exp{(Q)}.
\end{eqnarray} 
Note that $\Omega^2$ has a double property, in one had, it can be written as a purely a quantum mechanical quantity and on the other hand it has a definite geometrical meaning as a conformal factor which can be written in terms of a Weyl scalar field. Substituting the expression of $\Omega^2$ in Eq.~\ref{ExoticWEq}, one can write the equation in terms of the quantum potential $Q$.
 Hence the equation becomes,
 \begin{eqnarray}
{\Box{Q}+\frac{1}{4} {\nabla_{\sigma}Q \nabla^{\sigma}Q}}+\mbox{Alc}(x,y,z)=0
\end{eqnarray} 
Substituting the expression for $Q$ in terms of the quantum mechanical matter wave density, i.e using the following relation,
\begin{eqnarray} 
Q=\frac{\hbar^2}{m^2} {\Bigl(\frac{\Box{\sqrt{\rho}}}{\sqrt{\rho}}\Bigr)}.
\end{eqnarray} 
One can get,
\begin{eqnarray} 
\frac{\hbar^2}{m^2} \Box{{\Bigl(\frac{\Box{\sqrt{\rho}}}{\sqrt{\rho}}\Bigr)}} 
-\frac{1}{4}\frac{\hbar^4}{m^4} {\nabla_{\sigma}{\Bigl(\frac{\Box{\sqrt{\rho}}}{\sqrt{\rho}}\Bigr)} 
 \nabla^{\sigma}{\Bigl(\frac{\Box{\sqrt{\rho}}}{\sqrt{\rho}}\Bigr)} } \nonumber \\
 +\mbox{Alc}(x,y,z)=0
\end{eqnarray} 
Ignoring the higher order terms  i.e $O(\hbar^4)$ terms, one can obtain,

\begin{eqnarray} 
\Box{{\Bigl(\frac{\Box{\sqrt{\rho}}}{\sqrt{\rho}}\Bigr)}} 
 +\frac{m^2}{\hbar^2}\mbox{Alc}(x,y,z)=0 \label{4thWEq}
\end{eqnarray} 

Laplace-Beltrami operator on a curved space-time is given by,
\begin{eqnarray} 
\Box{\sqrt{\rho}}=\frac{1}{\sqrt{-g}}\partial_{\mu} \Bigl(\sqrt{-g}\,g^{\mu\nu}\partial_{\nu}{\sqrt{\rho}}\Bigr).
\end{eqnarray} 
Since  $\sqrt{-g}=1$ (for Alcubierre metric ), in our context,  the fourth order contribution $\Box{{\Bigl(\frac{\Box{\sqrt{\rho}}}{\sqrt{\rho}}\Bigr)}}$ becomes,
\begin{eqnarray} 
\Box{{\Bigl(\frac{\Box{\sqrt{\rho}}}{\sqrt{\rho}}\Bigr)}}=\partial_{\alpha} \Bigl(g^{\alpha\beta}\partial_{\beta}\Bigl(\frac{\partial_{\mu} \bigl(g^{\mu\nu}\partial_{\nu}{\sqrt{\rho}}\bigr)}{\sqrt{\rho}}\Bigr)\Bigr).
\end{eqnarray} 
Thus Eq.~\ref{4thWEq} becomes,
\begin{eqnarray} 
\partial_{\alpha} \Bigl(g^{\alpha\beta}\partial_{\beta}\Bigl(\frac{\partial_{\mu} \bigl(g^{\mu\nu}\partial_{\nu}{\sqrt{\rho}}\bigr)}{\sqrt{\rho}}\Bigr)\Bigr)
+\frac{m^2}{\hbar^2}\mbox{Alc}(x,y,z)=0 \label{F4Weq}\nonumber \\
\end{eqnarray} 
Expanding this equation in terms of the components of the metric tensor (given in Eq.\ref{Alcgdown}), one is lead to the expression containing the fourth order space-time derivative of $\sqrt{\rho}$ with other terms involving the Alcubierre shape function $f(r_{s})$ and the velocity of the warp bubble $v_{s}$. 

 It is difficult to find the analytical solutions of these kinds of fourth order differential equations but one can take very simplified toy versions of the wave equation in order to have an idea of the new physics emerging out from the equation presented here. 
The main achievement we have made in Eq.~\ref{F4Weq} is mapping the problem of the exotic matter into finding a suitable quantum density for the matter field as a solution to the fourth order partial differential equation on curved space-time. If the square root of the quantum density ($\sqrt{\rho}$) satisfy the wave equation given in Eq.~\ref{F4Weq}, it is equivalent to having an exotic matter in usual general relativity. Thus, one needs to expand the idea of having exotic matter into the determination of proper quantum mechanical density. Hence the gravitational problem is mapped into a purely quantum mechanical problem. In the Bohmian Quantum Gravity sense, these quantum mechanical density will result into a conformal wave equation.  

Now let us assume a simple fourth order wave equation where the space-time is flat and the quantum density is confined along the $x$-direction and instead of $\mbox{Alc}(x,y,z)$,  we have  just a constant $\beta^4$. Hence we can present a toy equation as follows,
\begin{eqnarray} 
\partial_{t}^4 \sqrt{\rho(t,x)}- \partial_{x}^4 \sqrt{\rho(t,x)} +\beta^4 \sqrt{\rho(t,x)}=0 
\end{eqnarray} 
By the method of separation of variables,  it is possible to find interesting solutions to the quantum density.  Apart from the usual trigonometric functions, hyperbolic functions also appear as the solutions which is not allowed in usual quantum theory, since the wave-equation is only second order. But for this toy equation, one can see that $\sqrt{\rho(t)}=c_{1} \cos(\gamma\, t)+c_{2} \sin(\gamma\, t)+c_{3} \cosh(\gamma\, t)+c_{4} \sinh(\gamma\, t)$, where $c_{1},c_{2},c_{3},c_{4}$ and $\gamma$ are constants. Similarly one can solve the spatial part of the quantum density i.e,  $\sqrt{\rho(x)}=d_{1} \cos(\delta\, x)+d_{2} \sin(\delta\, x)+d_{3} \cosh(\delta\, x)+d_{4} \sinh(\delta\, x)$, where $d_{1},d_{2},d_{3},d_{4}$ and $\delta=\sqrt[4]{\beta^4+\gamma^4}$ are constants. The total quantum density is given by $\sqrt{\rho(t,x)}=\sqrt{\rho(t)} \sqrt{\rho(x)}$. Hence the final solution consists of the product of trigonometric and hyperbolic functions. The presence of the hyperbolic terms arises from the fourth order derivative of the toy equation and these are extra solutions only appear from the conformal waves appearing in the Bohmian quantum gravity.  Trignometric solutions are already there in usual quantum theory but the hyperbolic solutions might be bringing the new physics here, due to the fact that they are the extra solutions appearing in the theory. Exotic matter that we perceive in usual General Relativity is a simple effect arises from the complicated  conformal wave equations presented  here (see Eq.~\ref{F4Weq}).
\section{Conclusions}
We have analyzed the Alcubierre warp drive in the Bohmian Quantum Gravity context. The exotic matter problem is mapped into a 4th order quantum mechanical density equation involving  Alcubeirre shape function. It is found that the fourth order wave equation in space-time can contribute to  extra solutions of  quantum density which are  usually not present in the  quantum theory since usual quantum theory is second order in nature.  We have taken a simple toy model of the partial differential equation and found that  hyperbolic functions of quantum density can appear which may contribute to the new physics appearing in the theory. If one can find a viable distribution of quantum mechanical density which obeys the wave equation given in Eq.~\ref{F4Weq}, we will resolve our problem of exotic matter for Alcubierre warp drive in Bohmian Quantum Gravity frame-work. The exact solution of quantum density can be evaluated in further research works.

%

\end{document}